\documentclass[conference]{IEEEtran}
\IEEEoverridecommandlockouts
\usepackage{cite}
\usepackage{amsmath,amssymb,amsfonts}
\usepackage{algorithmic}
\usepackage{graphicx}
\usepackage{textcomp}
\usepackage{xcolor}
\def\BibTeX{{\rm B\kern-.05em{\sc i\kern-.025em b}\kern-.08em
    T\kern-.1667em\lower.7ex\hbox{E}\kern-.125emX}}
\begin{document}

\title{Quantifying Fish School Fragmentation under Predation Using Stochastic Differential Equations \\

}

\author{\IEEEauthorblockN{Junyi Qi}
\IEEEauthorblockA{\textit{Graduate School of Bioresource and Bioenvironmental} \\
\textit{ Sciences, Kyushu University}\\
744 Motooka, Nishi Ward, Fukuoka 819-0395, Japan \\
qi.junyi.198@s.kyushu-u.ac.jp}
~\\
\and
\IEEEauthorblockN{Thorkil Casse}
\IEEEauthorblockA{\textit{Department of Social Sciences and Business} \\
\textit{Roskilde University, Denmark}\\
\textit{Faculty of Agriculture, Kyushu University, Japan} \\
744 Motooka, Nishi Ward, Fukuoka 819-0395, Japan \\
casse@ruc.dk }
\and
\IEEEauthorblockN{Masayoshi Harada}
\IEEEauthorblockA{\textit{Faculty of Agriculture, Kyushu University} \\
744 Motooka, Nishi Ward, Fukuoka 819-0395, Japan \\
mharada@bpes.kyushu-u.ac.jp}
\and
\IEEEauthorblockN{Linh Thi Hoai Nguyen}
\IEEEauthorblockA{\textit{International Institute for Carbon-Neutral Energy Research} \\
\textit{Kyushu University}\\
744 Motooka, Nishi Ward, Fukuoka 819-0395, Japan \\
linh@i2cner.kyushu-u.ac.jp }
\and
\IEEEauthorblockN{Ton Viet Ta*}
\IEEEauthorblockA{\textit{Faculty of Agriculture, Kyushu University} \\
744 Motooka, Nishi Ward, Fukuoka 819-0395, Japan \\
tavietton@agr.kyushu-u.ac.jp}
*Corresponding author
}

\maketitle

\begin{abstract}
This study builds upon our previously proposed stochastic differential equation (SDE)-based model to further investigate fish school fragmentation under predation. Specifically, we explore structural dynamics by incorporating graph-theoretic metrics—namely, the number of connected components—to quantify changes in prey school organization. Two quantitative indicators, first split time and final component count, are introduced to assess the timing and extent of group disintegration. Sensitivity analyses are performed on key parameters  to evaluate their influence on group stability under nearest attack and center attack strategies. We independently examine the effect of environmental noise on fish school cohesion. Simulation results show that parameter changes impact fish school fragmentation differently under the two predation strategies. High environmental noise also makes it difficult for the school to stay cohesive. This framework provides a structured and quantitative basis for assessing how fish schools respond to different predation strategies and environmental noise levels.

\end{abstract}


\section{Introduction}
Fish often form schools as a collective defense to reduce predation risk, with school structure playing a critical role in threat response. The organization of fish schools has been a focus of empirical and theoretical studies \cite{b1, b2, b3}. Under predatory pressure, schools may fragment into multiple clusters, and understanding these mechanisms is essential for comprehending group-level escape behavior.

Previous studies have developed particle-based models simulating schooling behavior via attraction and repulsion forces. Oboshi et al. \cite{b4} introduced a rule-based model based on nearest-neighbor distances, while Olfati-Saber \cite{b5} and D'Orsogna et al. \cite{b6} proposed differential equation models using attractive-repulsive forces. Vicsek et al. \cite{b9} designed a simple difference model where particles move with constant speed, updating direction based on neighbor alignment plus stochastic perturbations. Ta et al. \cite{b7.0,b7,b7.5,b7.9} studied noise-induced school aggregation using an ODE framework. In our previous work \cite{b8}, we proposed a stochastic differential equation (SDE)-based model to describe predator-prey dynamics. All the behavioral patterns observed in our simulations correspond to empirically reported phenomena and were documented in detail in our previous work.


In this study, we extend our previous SDE-based model by introducing graph-theoretic metrics, particularly the number of connected components, to quantify topological changes in school structure over time. Two key indicators—first split time and final component count—are proposed to evaluate the timing and extent of fragmentation. Parameter sensitivity analyses are conducted under two predation strategies: nearest attack and center attack. Additionally, we perform independent simulations to examine the effects of environmental noise on school cohesion.


The model incorporates intra-school (prey-prey, predator-predator) and inter-school (predator-prey) interactions, governed by attraction-repulsion forces, velocity alignment, and environmental noise. Prey individuals interact through attraction and repulsion, while predators target either the nearest prey or the prey centroid depending on the hunting strategy.


The complete formulation of the prey and predator dynamics, from our previous work, is given in Equation~\eqref{eq:predation}. It consists of two major components: prey school dynamics and predator school dynamics. Prey dynamics include attraction-repulsion, velocity alignment, and stochastic environmental noise, while predator dynamics incorporate internal coordination and predator-prey interactions based on hunting tactics. Positions and velocities of the $i$-th prey and $k$-th predator in $\mathbb{R}^3$ are denoted by $x_i, v_i$ and $y_k, u_k$, respectively. The collective states are $\mathbf{X}, \mathbf{V}, \mathbf{Y}, \mathbf{U}$, with $N$ and $M$ representing the numbers of prey and predators. The Euclidean norm $\|\cdot\|$ measures distances. Interaction strengths are controlled by parameters: $\alpha$ and $\beta$ (prey-prey attraction-repulsion and velocity alignment), and $\delta$ (predator-prey escape interactions). Biologically, as $\alpha$ increases, behavior shifts from loose scattering to tighter cohesion. Higher $\beta$ enhances synchronized motion, but excessively high values may suppress cohesion and lead to disintegration. $\delta$ governs prey sensitivity to predators—low values result in delayed responses, while higher values induce stronger escape reactions and early fragmentation. The exponents $p$ and $q$ regulate the decay of prey forces, while $r, R, R_1$ define characteristic interaction distances. The exponents $\theta$ and $\theta_1$ govern interaction decay rates. A small constant $\epsilon$ prevents singularities as $\|x_i - x_j\| \to 0$. Stochastic terms $dw_i(t)$ and $dw_k(t)$ simulate environmental noise, with strengths $\sigma_i$ and $\sigma_k$.

\begin{equation}
\left\{
\begin{aligned}
dx_i &= v_i \, dt + \sigma_i \, dw_i(t), \quad i = 1, 2, \ldots, N, \\
dv_i &= \Bigg[ 
    - \alpha \sum_{\substack{j=1 \\ j \neq i}}^N 
    \left( \frac{r^p}{\|x_i - x_j\|^p + \epsilon} 
    - \frac{r^q}{\|x_i - x_j\|^q + \epsilon} \right) \\
&\quad 
    \times (x_i - x_j) - \beta \sum_{\substack{j=1 \\ j \neq i}}^N 
    \left( \frac{r^p}{\|x_i - x_j\|^p + \epsilon} \right. \\
&\quad 
    \left. + \frac{r^q}{\|x_i - x_j\|^q + \epsilon} \right)(v_i - v_j) \\
&\quad     
    + \delta \sum_{k=1}^M 
    \left( \frac{R_1^{\theta_1}}{\|x_i - y_k\|^{\theta_1} + \epsilon} \right)(x_i - y_k)\\
&\quad     
    + F_i(t, x_i, v_i) \Bigg] dt, \quad i = 1, 2, \ldots, N,\\
dy_k &= u_k \, dt + \sigma_k \, dw_k(t), \quad k = 1, 2, \ldots, M, \\
du_k &= \left[
    \delta \sum_{\substack{j=1 \\ j \neq k}}^M 
    \left( \frac{R^\theta}{\|y_k - y_j\|^\theta + \epsilon} \right)(y_k - y_j)\right.\\
&\quad
    \left. + F(\mathbf{X}, \mathbf{V}, \mathbf{Y}, \mathbf{U})
\right] dt, \quad k = 1, 2, \ldots, M,
\end{aligned}
\right.
\label{eq:predation}
\end{equation}

\begin{figure}[htbp]
\centering
\includegraphics[width=0.7\linewidth]{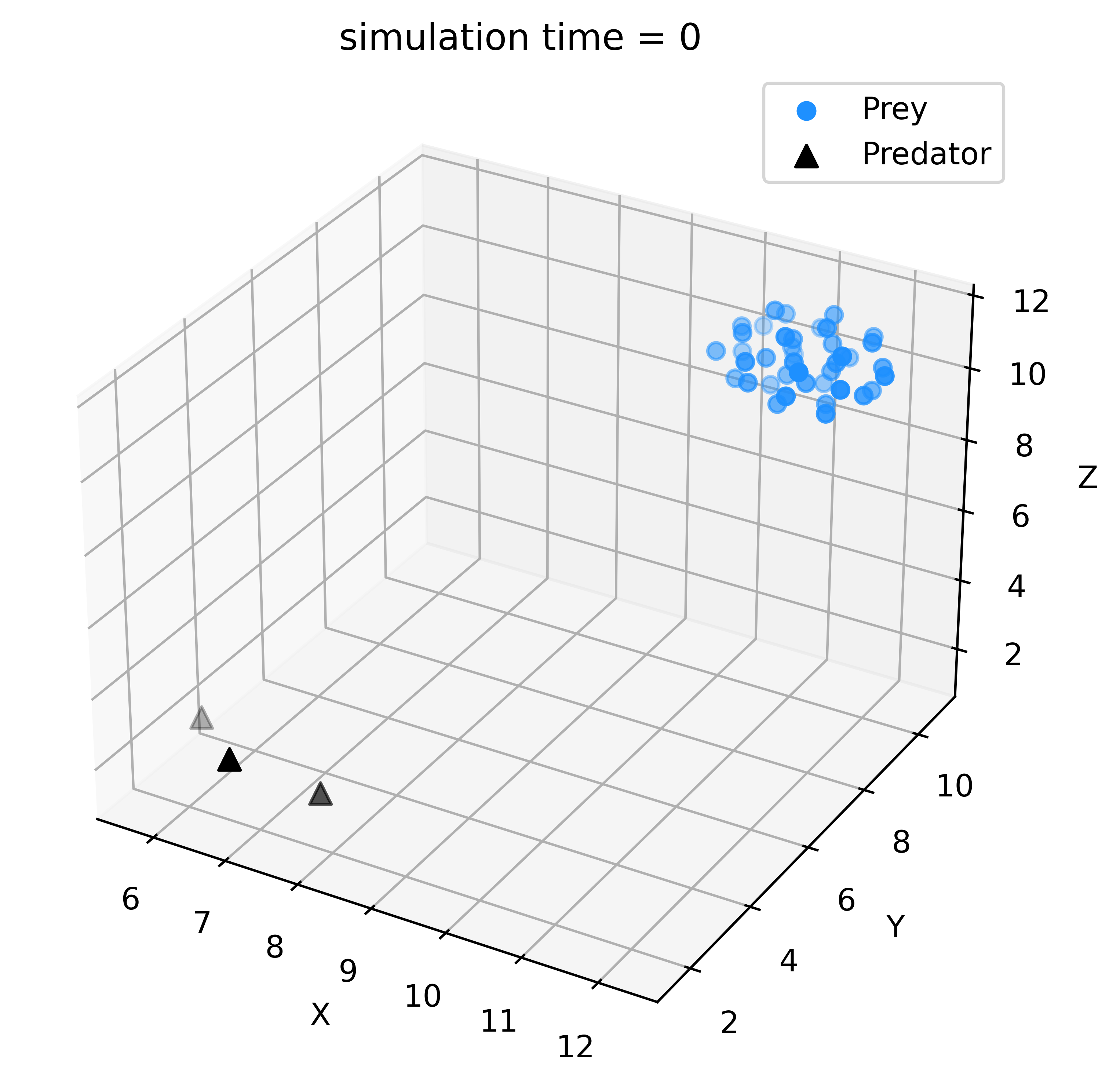}
\caption{The initial spatial configuration of prey and predator schools.}
\label{fig1}
\end{figure}

To initialize the simulation, we first generate separate cohesive prey and predator schools using the schooling model described in Equation~\eqref{eq:schooling}. Then, we compute the geometric center of each group and reposition them such that the distance between the two centers is fixed, while preserving the internal structure of each school. This spatial arrangement serves as the initial condition for the predator-prey interaction model given in Equation~\eqref{eq:predation}. Figure~\ref{fig1} illustrates this initial configuration.

\begin{equation}
\left\{
\begin{aligned}
dx_i &= v_i \, dt + \sigma_i \, dw_i, \quad i = 1, 2, \ldots, N, \\
dv_i &= \Bigg[
    -\alpha_1 \sum_{\substack{j=1 \\ j \neq i}}^N 
    \left( \frac{1}{\|x_i - x_j\|^p + \epsilon} 
    - \frac{\gamma}{\|x_i - x_j\|^q + \epsilon} \right) \\
&\quad  \times  (x_i - x_j) 
    - \beta_1 \sum_{\substack{j=1 \\ j \neq i}}^N 
    \left( \frac{1}{\|x_i - x_j\|^p + \epsilon} \right. \\
&\quad\quad 
    \left. + \frac{\gamma}{\|x_i - x_j\|^q + \epsilon} \right)  (v_i - v_j) 
    + F_i(t, x_i, v_i) \Bigg] dt, \\
&\quad i = 1, 2, \ldots, N.
\end{aligned}
\right.
\label{eq:schooling}
\end{equation}

In our previously proposed model, we consider two primary hunting strategies used by predators: 
\begin{itemize}
\item In the nearest attack strategy, predators pursue the closest prey individual. We define this strategy using the following force function:

\begin{equation}
\begin{aligned}
F(\mathbf{X}, \mathbf{V}, \mathbf{Y}, \mathbf{U}) 
&= -\frac{1}{N} \sum_{j=1}^{N} \frac{R_2^{\theta_2}}{\|{y}_k - {x}_j\|^{\theta_2} + \epsilon} \\
&\quad \cdot\left[ \gamma_1 ({y}_k - {x}_j) + \gamma_1 \gamma_2 ({u}_k - {v}_j) \right]
\end{aligned}
\label{eq:nearest_attack}
\end{equation}

\item In the center attack strategy, each predator targets the centroid of the prey school. The center is calculated as the mean position and velocity of all prey. Specifically, the mean position $x_c$ and mean velocity $v_c$ are defined as:

\begin{equation}
x_c = \frac{1}{N} \sum_{i=1}^{N} x_i, \quad 
v_c = \frac{1}{N} \sum_{i=1}^{N} v_i.
\end{equation}

We define this strategy using the following force function:

\begin{equation}
\begin{aligned}
F(\mathbf{X}, \mathbf{V}, \mathbf{Y}, \mathbf{U}) 
&= -\frac{R_2^{\theta_2}}{\|{y}_k - {x}_c\|^{\theta_2} + \epsilon} \\
&\quad \cdot \left[ \gamma_1 ({y}_k - {x}_c) + \gamma_1 \gamma_2 ({u}_k - {v}_c) \right]
\end{aligned}
\label{eq:center_attack}
\end{equation}

\end{itemize}

To assess the robustness of the model, we perform parameter sensitivity analyses under two distinct predation strategies: nearest attack and center attack. We also conduct an independent set of simulations to examine the role of environmental noise in affecting school cohesion.

This framework enables a quantitative understanding of how internal interaction parameters and external perturbations jointly shape the structural stability of fish schools. The results offer broader insights into the vulnerability of collective systems under varying ecological pressures.

\section{Effects of Parameters Including Noise}
We begin by establishing a baseline parameter setting to investigate how fish school fragmentation responds to variations in model parameters under predation. The default values are: $\alpha=4$, $\beta=0.1$, $p=4$, $q=5$, $r=1$, $\gamma_1=5$, $\gamma_2=0.2$, $R=2$, $\theta=2$, $R_1=2$, $\theta_1=1$, and $\delta=1.5$. Noise levels are fixed at $\sigma_i = 0.015$ and $\sigma_k = 0.016$. The prey school size is $N=40$, the predator school size is $M=3$, the time step is $\Delta t = 0.01$, and the total simulation time is 30. These values serve as the default configuration for all subsequent sensitivity tests. To evaluate the influence of a specific parameter, such as $\alpha$, we vary its value while keeping all others fixed.

\subsection{How the Prey School Structure Evolves}
We use the number of connected components as a graph-theoretic metric to characterize the degree of fragmentation in the fish school. A connected component is defined as a maximal subset of individuals in which any pair of individuals is connected—either directly or indirectly—by a path of individuals, each within a specified spatial distance of its neighbors. To evaluate prey school structure, we define connectivity based on spatial proximity: two individuals are considered connected if the distance between them is less than a threshold \( r^* = 1.2 \).

 In our model, \( r = 1 \) defines the critical distance for repulsion and attraction among prey individuals. Therefore, \( r^* = 1.2 \), slightly larger than this critical value, is selected as a practical cutoff to capture meaningful fragmentation events while avoiding over-sensitivity to small fluctuations. 
 
\begin{figure}[htbp]
\centering
\includegraphics[width=0.45\textwidth]{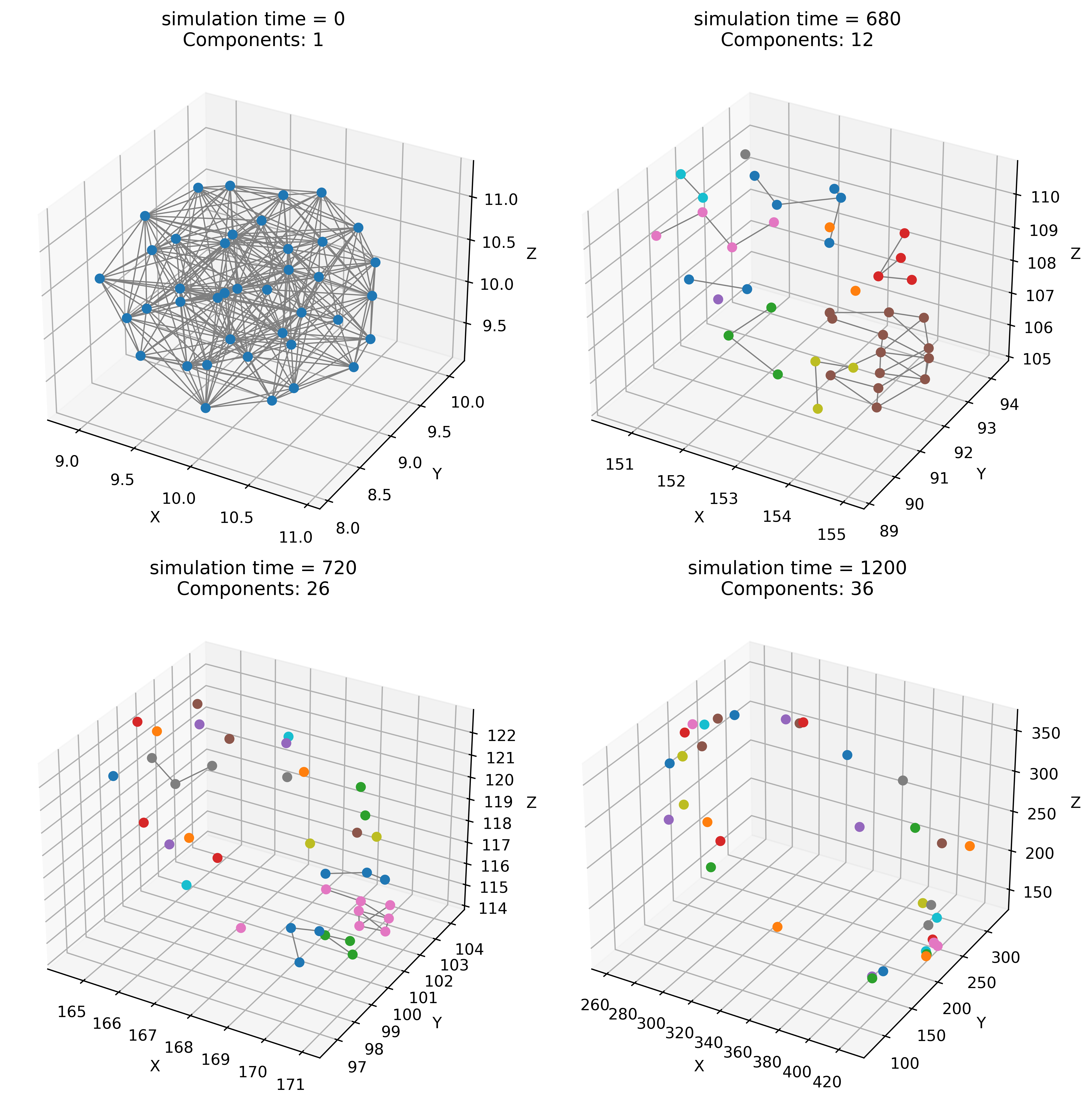}
\caption{Snapshots of the prey school at different times under nearest attack strategy. Individuals in the same connected component are shown in the same color.}
\label{fig2}
\end{figure}

\begin{figure}[htbp]
\centering
\includegraphics[width=0.45\textwidth]{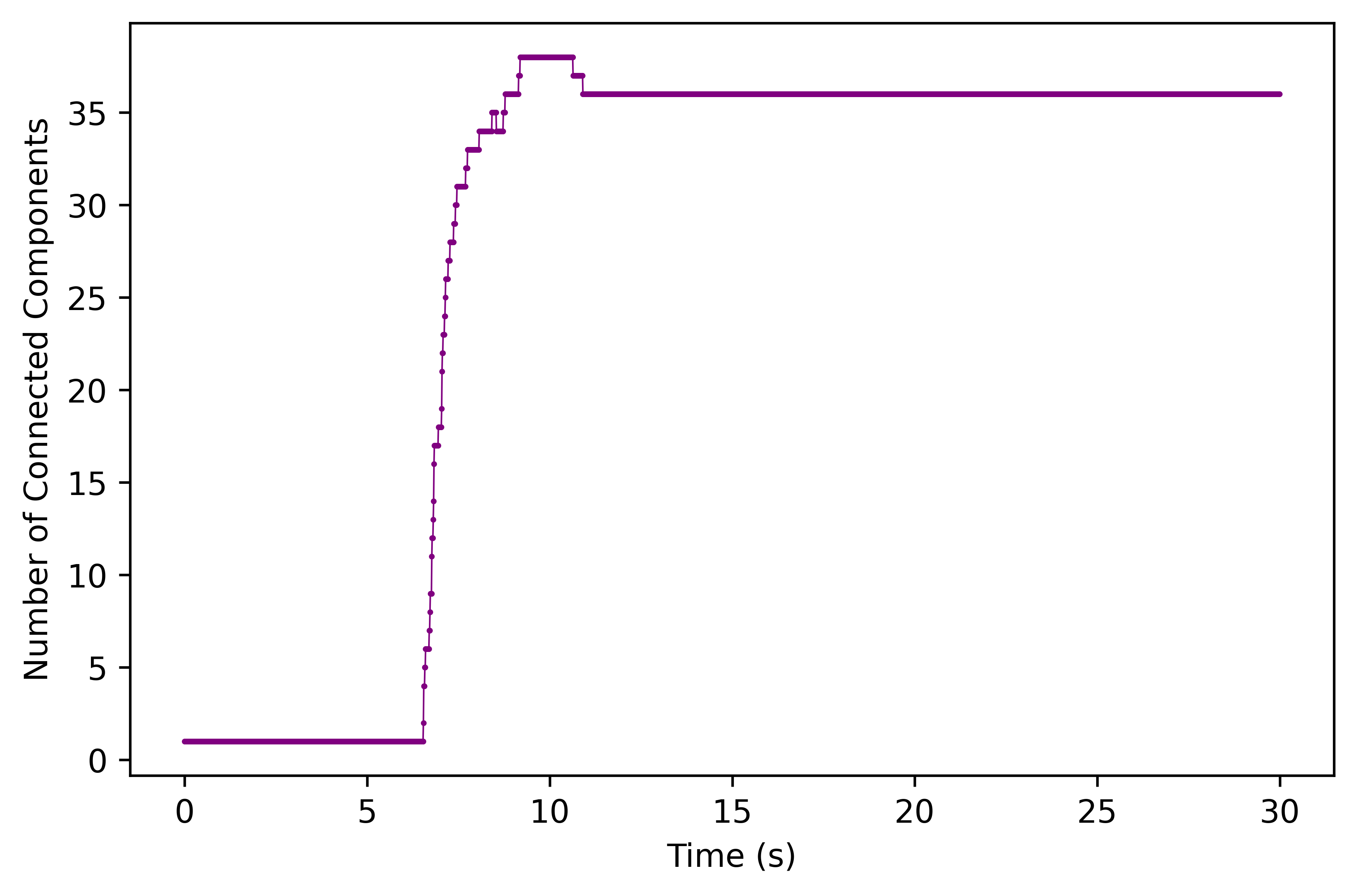}
\caption{Number of connected components over time under nearest attack strategy and the default parameter setting.}
\label{fig3}
\end{figure}

\begin{figure}[htbp]
\centering
\includegraphics[width=0.45\textwidth]{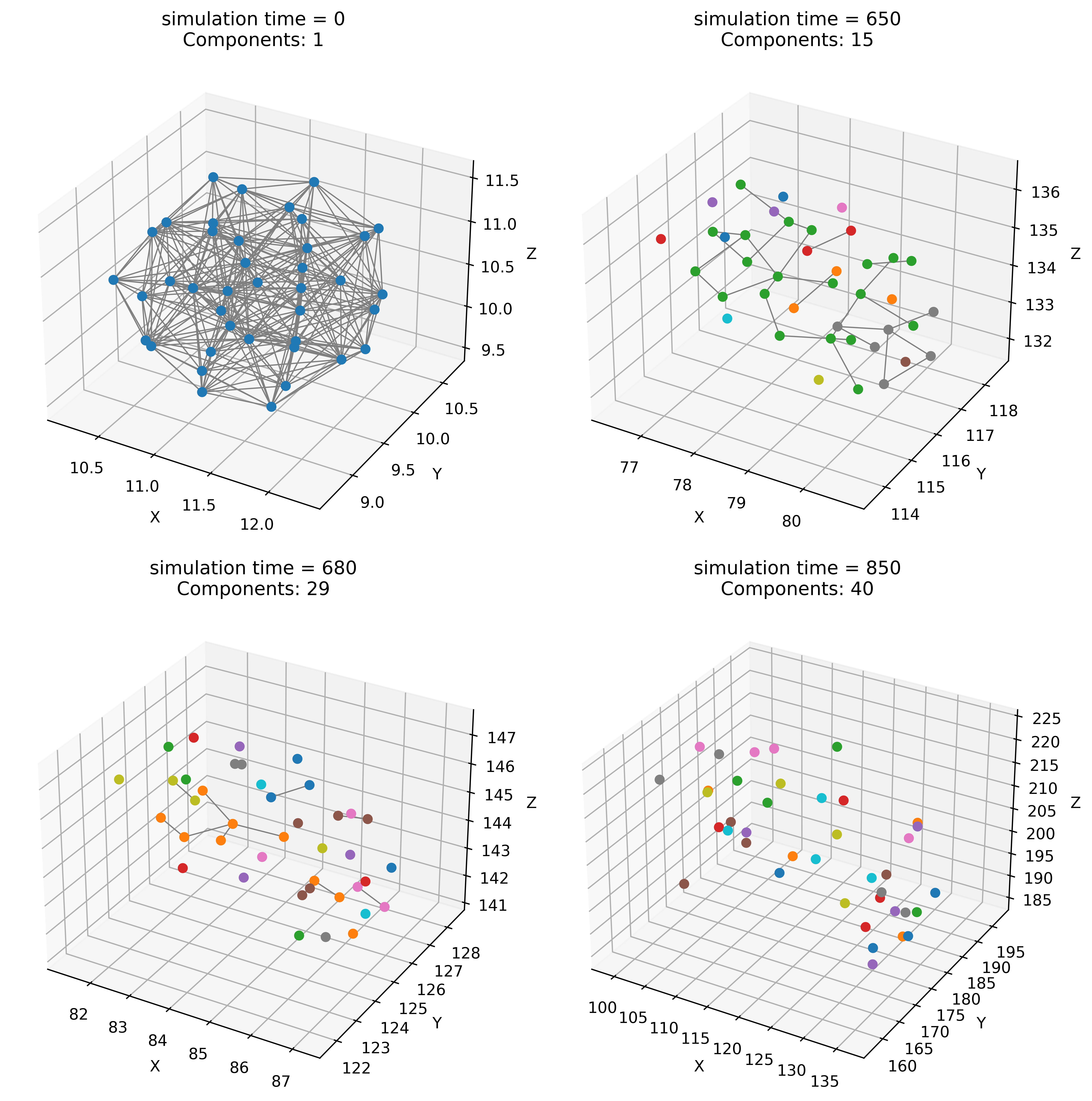}
\caption{Snapshots of the prey school at different times under center attack strategy. Individuals in the same connected component are shown in the same color.}
\label{fig4}
\end{figure}

\begin{figure}[htbp]
\centering
\includegraphics[width=0.45\textwidth]{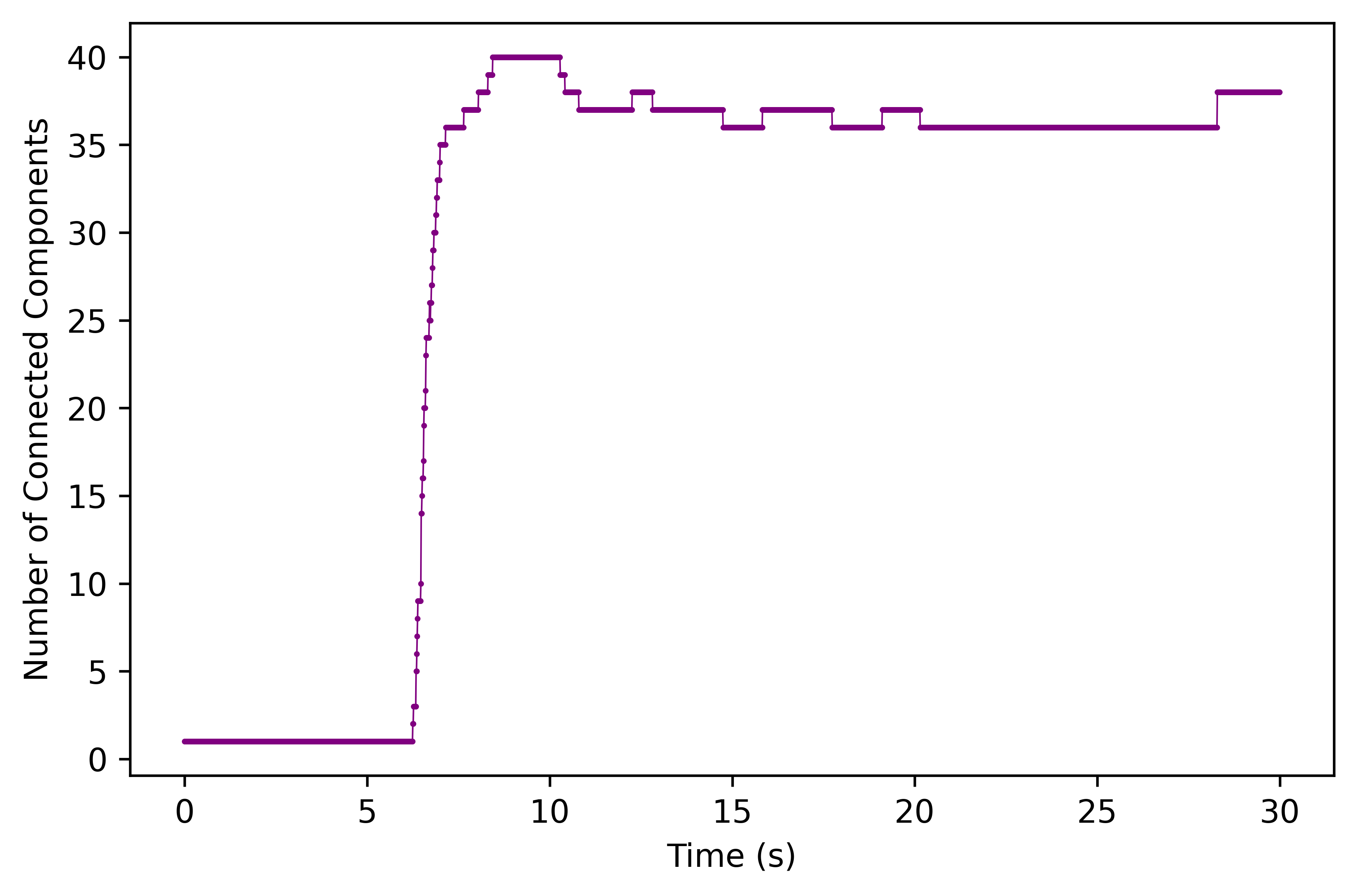}
\caption{Number of connected components over time under center attack strategy and the default parameter setting.}
\label{fig5}
\end{figure}

 Figure~\ref{fig2} and Figure~\ref{fig3} illustrate the changes in the number of connected components over time under nearest attack strategy, Figure~\ref{fig4} and Figure~\ref{fig5} illustrate the changes in the number of connected components over time under center attack strategy, based on the same initial parameter setting. In these 3D visualizations, the prey school initially starts as a single cohesive group and gradually splits into multiple disconnected clusters under predation pressure. 
 
 It is worth noting that the final number of connected components is slightly lower in the nearest attack case (36) than in the center attack case (40), suggesting stronger local cohesion among prey individuals under nearest attack. These visual and quantitative results indicate that the nearest attack strategy results in more cohesive small clusters, while the center attack strategy causes more dispersed disintegration. This observation supports findings by Demšar et al. \cite{b10}, who demonstrated that such center-focused tactics can effectively disrupt the structural integrity of fish schools.

\subsection{Sensitivity Analysis of Turning Parameters}

\begin{figure}[htbp]
\centering
\includegraphics[width=0.45\textwidth]{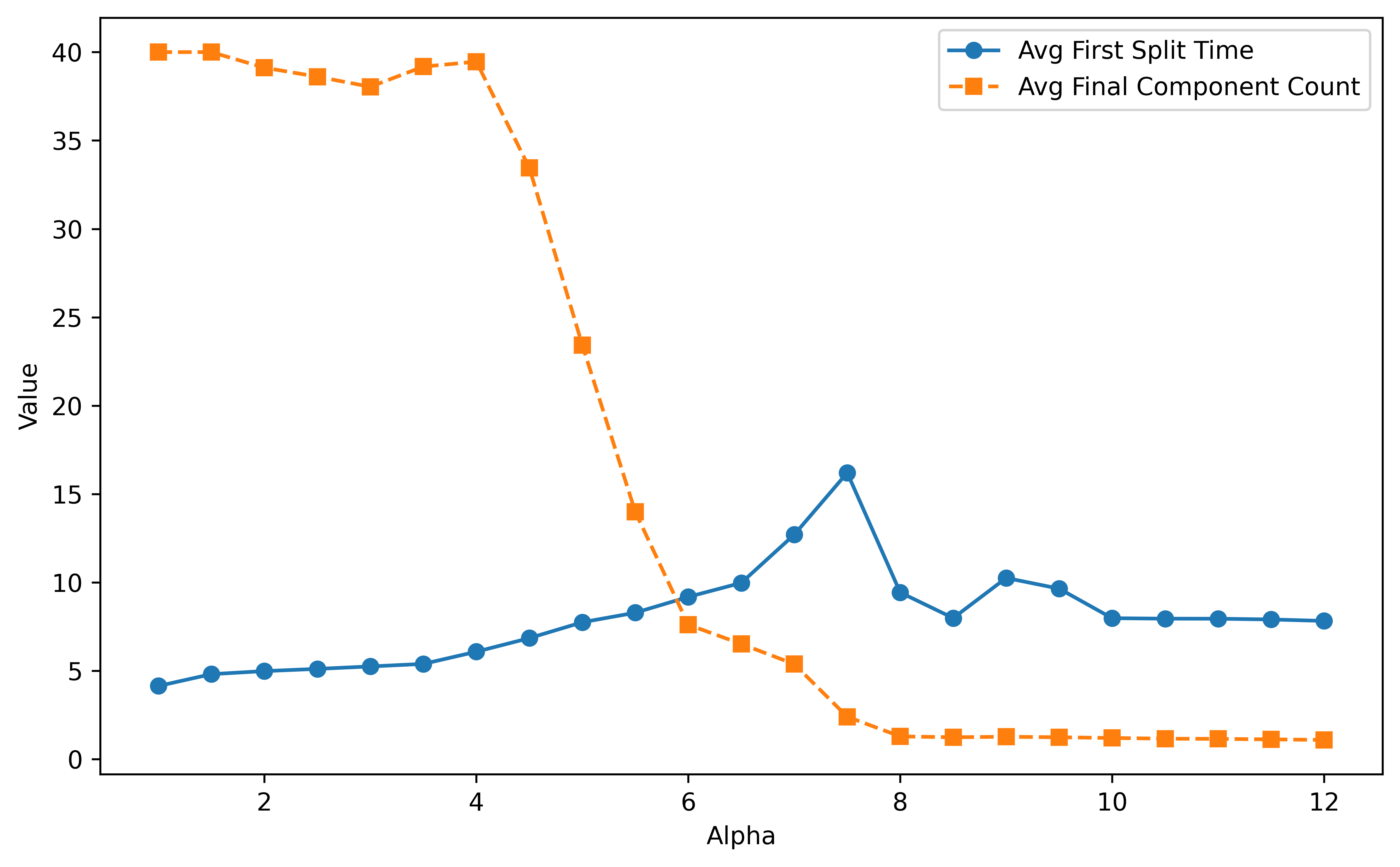}
\caption{Effect of $\alpha$ on first split time and final component count under nearest attack strategy.}
\label{fig:alpha_nearest}
\end{figure}

\begin{figure}[htbp]
\centering
\includegraphics[width=0.45\textwidth]{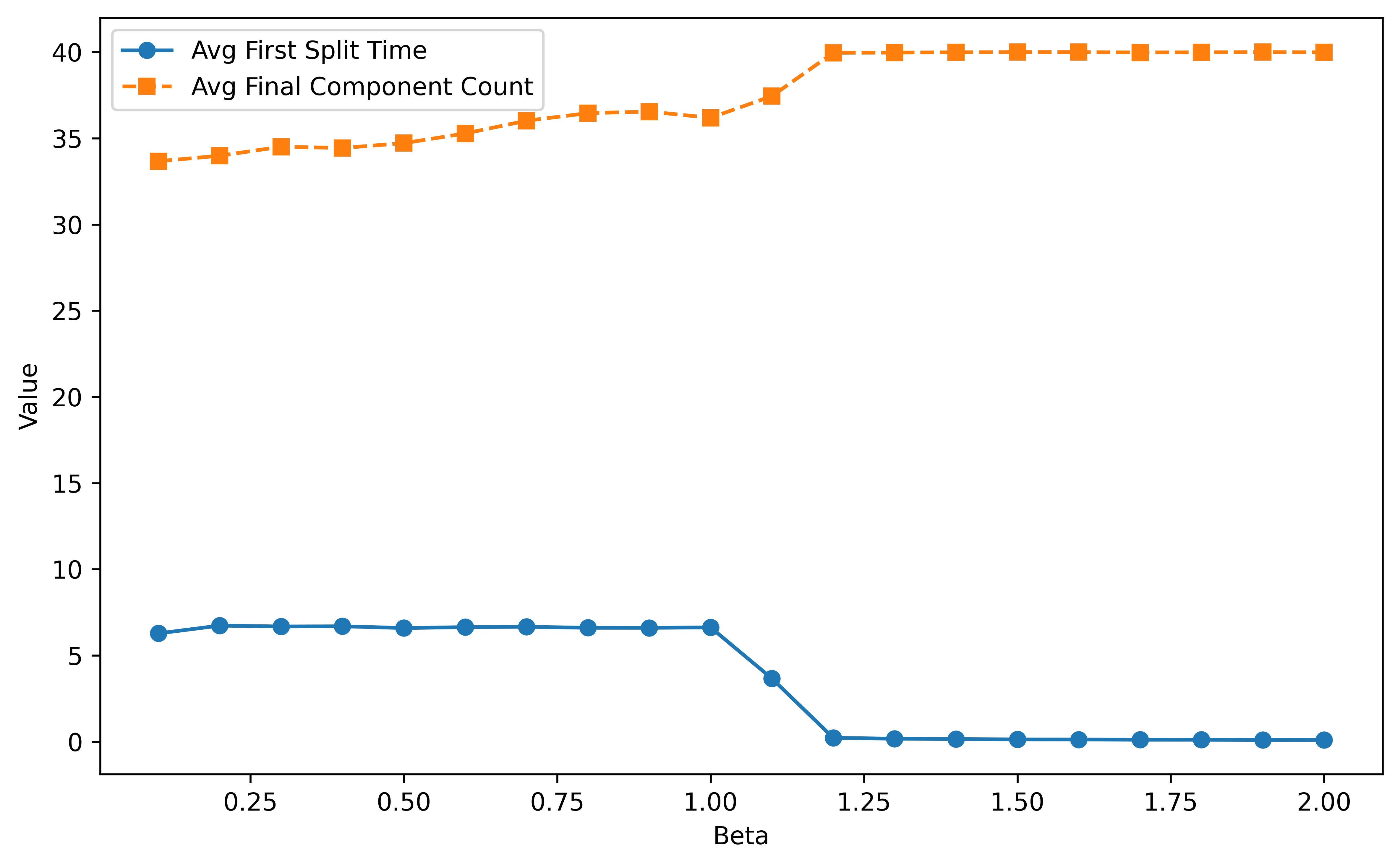}
\caption{Effect of $\beta$ on first split time and final component count under nearest attack strategy.}
\label{fig:beta_nearest}
\end{figure}

To investigate parameter sensitivity, we explore the effects of the key parameters $\alpha$ and $\beta$, which control the strength of attraction-repulsion and velocity alignment between prey individuals, respectively.  For each parameter value, we perform 100 simulation trials and compute the average first split time and final component count. This allows us to evaluate how these parameters influence the fragmentation dynamics of the fish school under different predation strategies. Figures~\ref{fig:alpha_nearest} and~\ref{fig:beta_nearest} present the effects of parameters $\alpha$ and $\beta$ on the first split time and final number of connected components under the nearest attack strategy.

Under nearest attack strategy, our simulations reveal that changes in the turning parameters $\alpha$ and $\beta$ significantly affect both the fragmentation timing and the final structure of the prey school. Increasing $\alpha$, which controls the strength of attraction and repulsion among individuals, enhances cohesion and delays fragmentation. Initially, low values of $\alpha$ result in early and severe fragmentation. As $\alpha$ increases, the prey school becomes more cohesive, eventually stabilizing into two compact clusters under predation pressure. In contrast, $\beta$ controls velocity alignment among prey individuals. Moderate values of $\beta$ enhance coordination and improve prey school cohesion. However, when $\beta$ becomes too large, it leads to full disintegration—resulting in 40 disconnected components. This result occurs because strong velocity alignment suppresses attraction forces, making it difficult for individuals to return to or maintain group structure. Moreover, high $\beta$ increases sensitivity to environmental noise, causing small perturbations to propagate rapidly and destabilize the school. Each individual follows local velocity, eventually leading to complete structural breakdown.

\begin{figure}[htbp]
\centering
\includegraphics[width=0.45\textwidth]{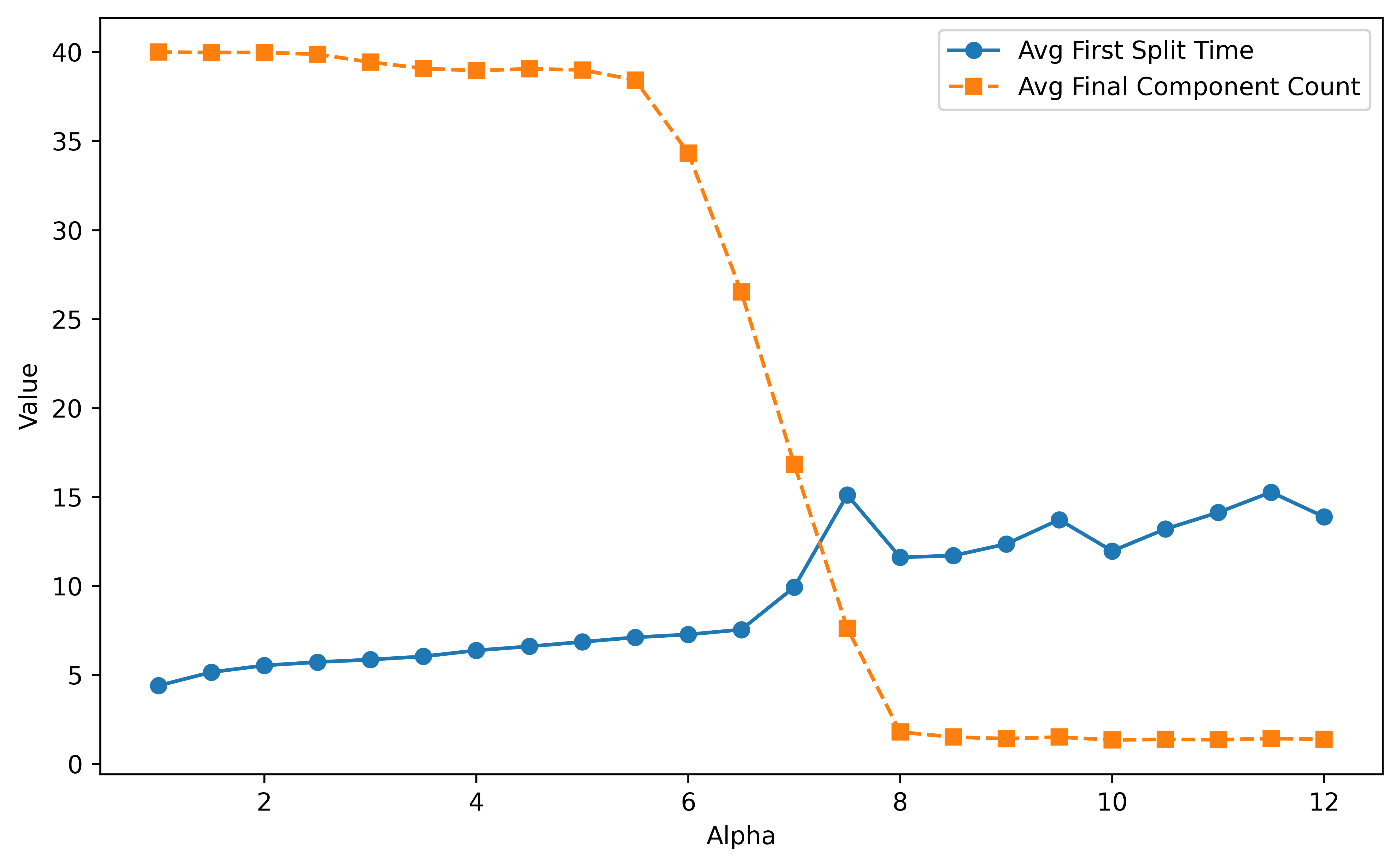}
\caption{Effect of $\alpha$ on first split time and final component count under center attack strategy.}
\label{fig:alpha_center}
\end{figure}

\begin{figure}[htbp]
\centering
\includegraphics[width=0.45\textwidth]{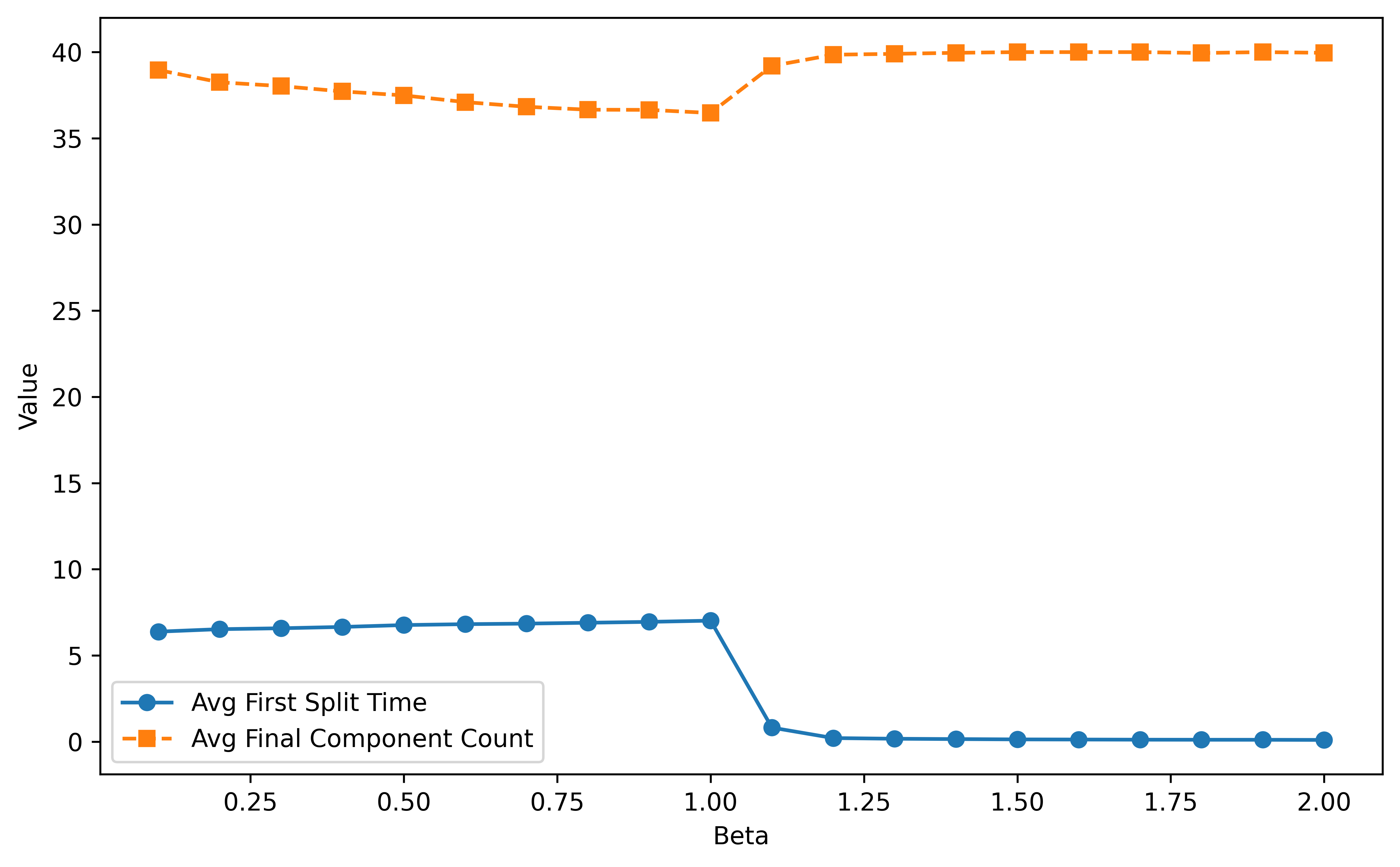}
\caption{Effect of $\beta$ on first split time and final component count under center attack strategy.}
\label{fig:beta_center}
\end{figure}

Figures~\ref{fig:alpha_center} and~\ref{fig:beta_center} present the effects of parameters $\alpha$ and $\beta$ on the first split time and final number of connected components under the center attack strategy.

Under the center attack strategy, the sensitivity trends of parameters $\alpha$ and $\beta$ remain generally consistent with those observed under the nearest attack strategy. As $\alpha$ increases, the average first split time gradually increases, indicating stronger group cohesion and delayed fragmentation. Meanwhile, the average final number of connected components decreases and approaches 1, suggesting that the prey school tends to remain as a single cohesive group rather than splitting apart. This reflects a balance between repulsion and cohesion at higher $\alpha$ values. For $\beta$, when it exceeds a certain threshold, the average first split time drops significantly, and the final component count rises to 40, indicating complete fragmentation. This can be attributed to the dominance of velocity alignment, which suppresses attractive forces and increases sensitivity to noise, leading to a structural breakdown of the school.

Under relatively high $\alpha$, the center attack strategy results in a higher degree of fragmentation compared to the nearest attack strategy. This reflects the stronger disruptive effect of targeting the centroid of the prey school, as the center attack strategy more effectively breaks its cohesion. This consistents with the findings of Demšar et al. \cite{b10}, who demonstrated that such tactics disrupt the school’s core structure and facilitate individual isolation by predators.

\begin{figure}[htbp]
\centering
\includegraphics[width=0.45\textwidth]{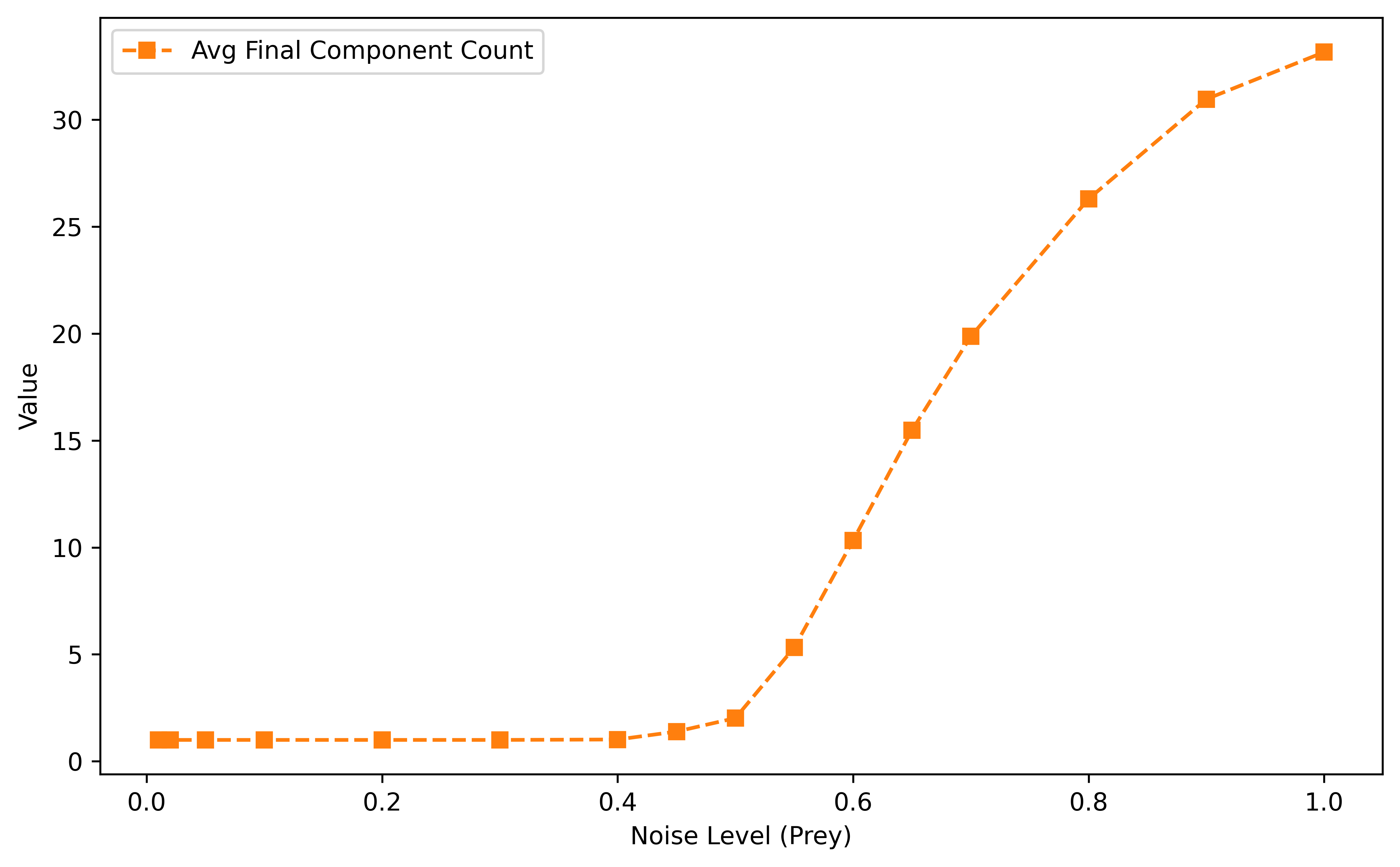}
\caption{Effect of prey noise level on the final number of connected components.}
\label{fig:noise_prey}
\end{figure}

We further examine how environmental noise affects the structural stability of prey schools. As shown in Figure~\ref{fig:noise_prey}, the final number of connected components increases sharply with higher prey noise levels. When the noise level is below 0.4, indicating that the prey school maintained cohesion. However, when the noise level exceeded this threshold, the prey school began to fragment rapidly, and the final number of components increased significantly. This result suggests that excessive environmental noise can disrupt prey cohesion.

\section{Conclusion}

In this study, we extended our previously developed SDE-based predator-prey model to investigate the fragmentation behavior of fish schools under predation. We used graph metrics like connected components to measure group structure and introduced two indicators—first split time and final component count—to assess cohesion.
Results show that hgher $\alpha$ improves cohesion, while overly large $\beta$ leads to full disintegration due to excessive alignment and noise sensitivity. Center attack strategies cause more disruption under high $\alpha$ than nearest attack. Additionally, high environmental noise can rapidly break school structure. These findings highlight how internal rules and external disturbances together affect collective behavior stability.



\vspace{12pt}


\begin{thebibliography}{00}
\bibitem{b1} A. D. Hartono, L. T. H. Nguyen, and T. V. Ta, ``A stochastic differential equation model for predator-avoidance fish schooling,'' Math. Biosci., vol. 367 (2024), 109112.

\bibitem{b2}T. V. Ta and L. T. H. Nguyen, ``A stochastic differential equation model for the foraging behavior of fish schools,'' Phys. Biol., vol. 15 (2018), 036007.

\bibitem{b3} S. Camazine, J. L. Deneubourg, N. R. Franks, J. Sneyd, G. Theraulaz, and E. Bonabeau, ``Self-Organization in Biological Systems,'' Princeton, NJ: Princeton University Press (2001).

\bibitem{b4} T. Oboshi, S. Kato, A. Mutoh, and H. Itoh, ``Collective or scattering: evolving schooling behaviors to escape from predator,'' Artificial Life, vol. VIII (2002), pp. 386--389.

\bibitem{b5} R. Olfati-Saber, ``Flocking for multi-agent dynamic systems: Algorithms and theory,'' IEEE Trans. Automat. Control, vol. 51(2006), pp. 401--420.

\bibitem{b6} M. R. D'Orsogna, Y. Chuang, A. Bertozzi, and L. Chayes, ``Self-propelled particles with soft-core interactions: patterns, stability and collapse,'' Phys. Rev. Lett., vol. 96 (2006), 104302.

\bibitem{b9} T. Vicsek, A. Czirók, E. Ben-Jacob, I. Cohen, and O. Shochet, ``Novel type of phase transition in a system of self-driven particles,'' Phys. Rev. Lett., vol. 75 (1995), pp. 1226--1229.

\bibitem{b7.0} T. Uchitane, T. V. Ta, and A. Yagi, ``An ordinary differential equation model for fish schooling,'' Sci. Math. Jpn., vol. 75 (2012), pp. 339--350.

\bibitem{b7} L. T. H. Nguyen, T. V. Ta, and A. Yagi, ``Quantitative investigations for ODE model describing fish schooling,'' Sci. Math. Jpn., vol. 77, No. 3 (2014), pp. 403--413, [Scientiae Mathematicae Japonicae Online, e-2014, pp. 97--107].

\bibitem{b7.5} T. V. Ta, L. T. H. Nguyen, and A. Yagi, ``Flocking and non-flocking behavior in a stochastic Cucker-Smale system,'' Anal. Appl., vol. 12 (2014), pp. 63--73.

\bibitem{b7.9} L. T. H. Nguyen, T. V. Ta, and A. Yagi, ``Obstacle avoiding patterns and cohesiveness of fish school,'' J. Theoret. Biol., vol. 406 (2016), pp. 116--123.



\bibitem{b8} J. Qi and T. V. Ta, ``Modeling predator-prey dynamics with stochastic differential equations: patterns of collective hunting and nonlinear predation effects,'' preprint submitted to a journal.


\bibitem{b10} J. Demšar, C. K. Hemelrijk, H. Hildenbrandt, and I. L. Bajec, ``Simulating predator attacks on schools: Evolving composite tactics,'' Ecological Modelling, vol. 304 (2015), pp. 22--33.


\end{thebibliography}
\end{document}